# Electrical, Optical and Structural Properties of Improved Transparent Conductive Oxide TCO Films

**Wahid Shams-Kolahi** [a,b] **and Nazir P. Kherani** [b]

a) Department of Electrical and Computer Engineering, University of Toronto, Toronto, Ontario M5S 3G4, Canada
b) PRESET Adress: PRISED Solar Inc., Toronto, ON M5V 3S5

**Abstract**
High Frequency (GW) Post Processing of Transparent Conductive Oxide, ITO, leads to a significant improvement of Optical Transmission and Electrical Conduction. This is also true in case of ITO films in multilayered structures. The improvements happen without any affecting the lower layers (other layers of the multilayered Device).
This Method is extremely selective and can process only the targeted film (layer).

## 1. Itrouction

There is still a continuing interest in the development of low temperature deposition of TCO films because of their importance in a wide range of applications such as photovoltaics, liquid crystal displays and photoconductors [1 to 4]. Indium tin oxide (ITO) is one of such materials. The requirement of ITO films for the above mentioned applications is to have films with high conductivity and high transparency in the visible range using various deposition methods of which the most widely used ones are thermal evaporation [5] and sputtering [6].

Here we present the results of our study of the high frequency (GW) annealing on the structural, electrical and optical properties of ITO films deposited at low temperature using rf-magnetron sputtering. We have observed a significant improvement in the physical properties of room temperature deposited ITO films (comparable to the commercially available films).

There are commercially available ITO Films with very high electrical and optical Properties; such as an average transmission over 90% in the range of wavelength of 400-900nm and sheet resistances less than 15 $\Omega/\square$, however these films are deposited at high temperatures (>350°C).
Since the amorphous silicon deposition for hetero-junction solar cells is being carried out at temperatures below 200°C, the ITO deposition is also restricted to this temperature. We have deposited the ITO films at room temperature which leads to ITO films with high resistivity and low transmitivity compared to commercially available films.
With Rapid Thermal Annealing (RTA) at a temperature bellow 200°C, we have ALSO observed a significant improvement in the physical properties of our ITO films.

## 2. Samples and Methodology

ITO films, with thickness of 100.00–150 nm range, were deposited on glass substrates using rf-magnetron sputtering. In the sputtering system we used a sintered ITO target with an $In_2O_3:SnO_2$ composition of 90:10 wt. %. The substrate temperature during deposition was maintained at low value. The sputtering deposition was carried out in a pure argon atmosphere at a pressure from $\sim10^{-5}$ to $\sim10^{-3}$ torr and the sputtering power at ~150 W. Before the sputtering deposition, pre-sputtering is always carried out under the same condition as the deposition.

The surface morphology of the ITO films was studied by an Atomic Force Microscope (AFM) and scanning electron microscope (SEM). X-ray diffraction (XRD) measurements were performed using high resolution X-ray diffractometer system.

XRD measurements: The samples were run on a D8 Discovery Diffractometer system with Cu-kα source operating at 45 kV/45 mA. The system is equipped with 2D - proportional area detector (GADDS). The experimental data were collected on a single frame at 1200 s exposure that covered the range of $21^\circ$ – $39.5^\circ$ (2-theta). The obtained 2D diffraction image was then integrated in order to obtain standard, I vs. 2-theta, diffraction pattern. The data were processed by various Bruker AXS data processing software including Eva™ 8.0 and Topas™ v. 2.1 (for profile fitting and Rietvled analysis).

The sheet resistance of the samples was measured with a four-point probe and the resistivity of the film was calculated. The sheet resistance was calculated by simple relation $R_S = \rho/t$, where, $R_S$ is the sheet resistant, $\rho$ is the density and $t$ is the thickness.

The optical transmittance of the films was measured as the transmittance ratio of a film coated substrate relative to an uncoated substrate by UV Spectrophotometer .

## 3. Results and discussions

Fig. 1 shows the X-ray diffraction spectra of ITO films deposited on glass substrates. X-ray diffraction analysis indicated that the deposited films were polycrystalline. It can be seen that the ITO films deposited on glass substrate exhibit a preferred orientation. It may be pointed out that the preferred orientation for ITO films not only depends on the deposition conditions but also depends on the substrate material.

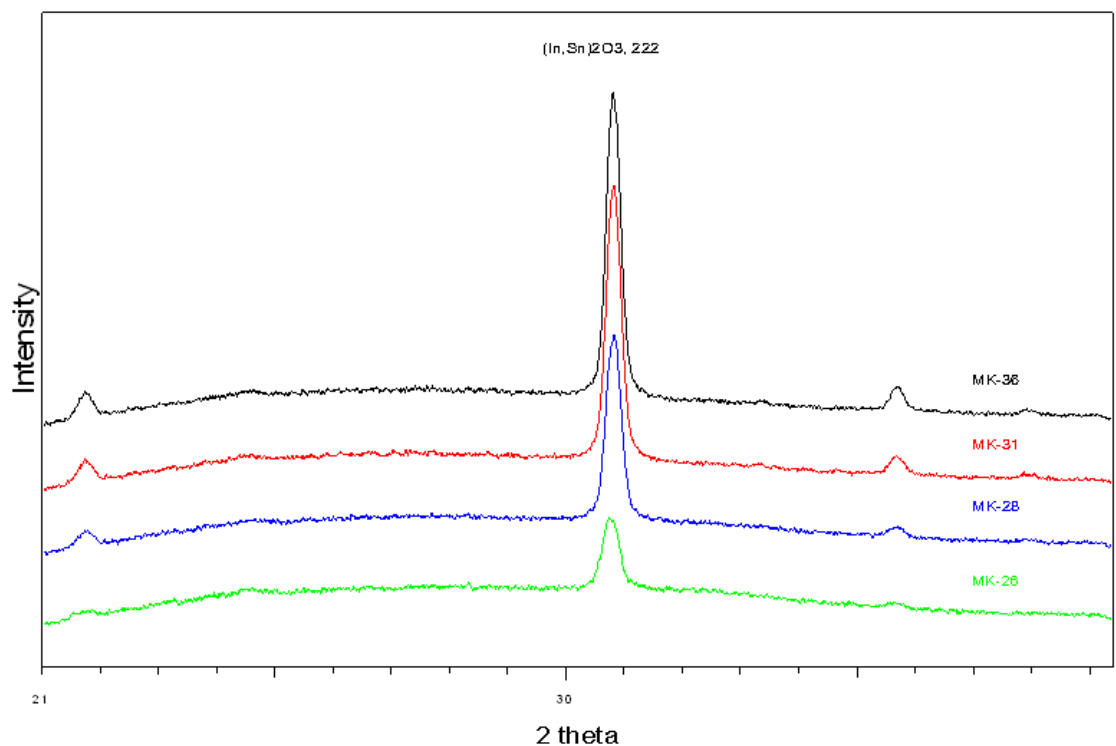

Fig. 1: PXRD patterns of thin film samples: MIK-26. MIK-28, MIK-31 and MIK-36

As we can understand it from figure 1, the crystalline nature becomes dominant with increase of the treatment of the layer which causes a significant increase of the intensity of the strongest 222 reflection, while the amorphous component visibly decreases. It is worth to note that the crystalline phase shows a random distribution of its diffraction crystallites that corresponds to a powdered sample.

**Electrical Properties**

The sheet resistance of ITO films is listed in Table1. It is found that ….

| HF Treatment (Sec.) | Sheet Res. (Glass) | Sheet Res. (Si) |
|---|---|---|
| 0 | 98 | 61.6 |
| 10 | 37.8 | 32.5 |
| 20 | 28.3 | 24.9 |
| 30 | 24.1 | 26.7 |
| 40 | 23.2 | 22.5 |
| 50 | 22.9 | 21.5 |
| 60 | 22.7 | 22.6 |
| 70 | 20 | 21.3 |
| 80 | 22.2 | 20.9 |
| 90 | 22.1 | 23.2 |
| 100 | 21.9 | 22.6 |

Table 1: Ddependency of the electrical resistivity on the HF-treatment of ITO films

The sheet resistance of ITO film decreases with time. There is a significant difference between sheet resistances before treatment, however the final resistances are almost equal.

**OPTICAL PROPERTIES**

The optical transmission of ITO films improves in a significant manner. A short HF treatment of 10 seconds increases the transmission more than 20% I higher photon energy regions, as show in Figure 2.

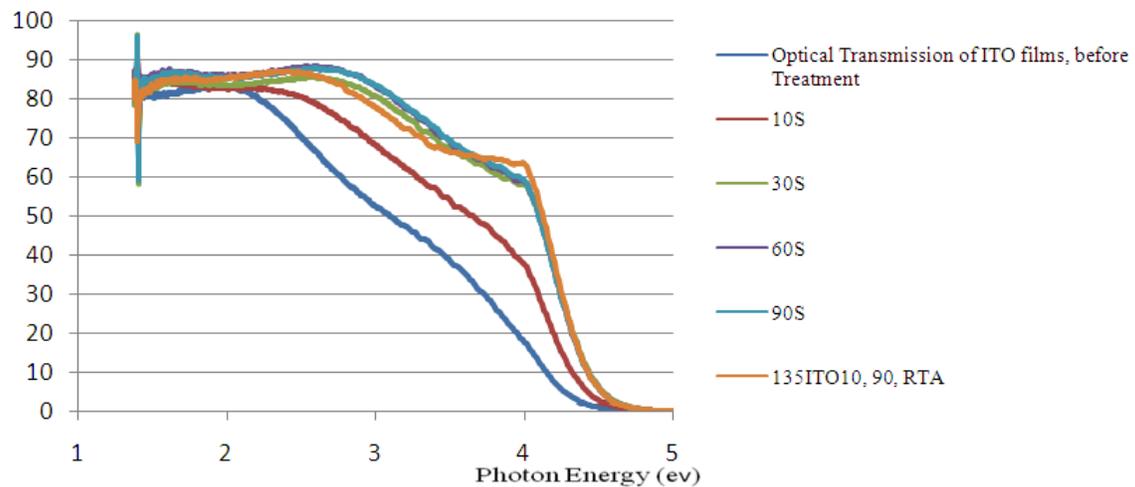

Fig.2 – Optical transmittance, T, as a function of the light wavelength, $\lambda$, for ITO thin films with various thicknesses deposited on PET substrates

## CONCLUSION

In conclusion, the surface morphology, the sheet resistance, the resistivity and the optical transmittance of HF treated ITO films show significant improvement of the film quality. This extremely short time post-process can be used for the performance of devices such as solar modules which suffer from low temperature deposition of their ITO films. The selective process possibility here allows us to improve the physical properties of sandwiched layers.

**Acknowledgments**

The authors are grateful to Dr. D. Yeghikian for valuable support on sample preparation and dr. S. Petrov for XRD analysis.

**References**
[1] P. P. Deimel, B. B. Heimhofer, G. Krotz, H. J. Lilienhof, J. Wind, G. Muller, and E. Voges, IEEE Photonic Technol. Lett. 2, 499 (1990).
[2] T. J. Coutts, X. Li, W. Wanlass, K. A. Emery, and T. A. Cessert, IEEE Electron. Lett. 26, 660 (1990).
[3] S. Honda, M. Watamori, and K. Oura, Thin Solid Films 281/282, 206 (1996).
[4] R. H. Bube, A. L. Fahrenbruch, R. Sinclair, T. S. Anthony, C. Fortman, C. T. Lee, T. Thorpe, and T. Yamashita, IEEE Trans. Electron Devices 31, 528 (1989).
[5] J. L. Yao, S. Hao, and J. S. Wilkinson, Thin Solid Films 189, 227 (1990).
[6] R. N. Joshi, V. P. Singh, and J. C. McClure, Thin Solid Films 257, 32 (1995).